\documentclass[twocolumn,floatfix,pra,showpacs]{revtex4}
\usepackage{amssymb}
\usepackage{amsmath}
\usepackage{color}
\usepackage{psfrag}
\usepackage{epsfig}
\usepackage{bbm}
\usepackage{bm}
\newcommand{\ket}[1]{| #1 \rangle}

\newcommand{\rb}[1]{\left( #1 \right)}

\newcommand{\ew}[1]{\langle #1 \rangle}
\newcommand{\beq}{\begin{eqnarray}}
\newcommand{\eeq}{\end{eqnarray}}

\newcommand{\svec}{\mbox{\boldmath$\sigma$}}

\newcommand{\sfrac}[2]{\begin{array}{c}\frac{#1}{#2}\end{array}}

\begin{document}
\title{Optically-controlled logic gates for two spin
  qubits in vertically-coupled quantum dots}
\author{C.~Emary\footnote{
Present address:
Institut f\"ur Theoretische Physik,
TU Berlin, 
Hardenbergstr. 36,
D-10623 BERLIN.}
 and L.~J.~Sham}
\affiliation{Department of Physics, University of California San
Diego, La Jolla, CA 92093, U.S.A.}
\date{\today}
\begin{abstract}
We describe an interaction mechanism between electron spins in a
vertically-stacked double quantum dot that can be used for
controlled two-qubit operations.
This interaction is mediated by excitons confined within, and
delocalized over, the double dot.
We show that gates equivalent to the $\sqrt{\mathrm{SWAP}}$ gate can
be obtained in times much less than the exciton relaxation time and
that the negative effects of hole-mixing and spontaneous emission do
not seriously affect these results.
\end{abstract}
\pacs{78.67.Hc, 03.67.Lx}
\maketitle


The spin of an electron confined in a quantum dot (QD) is one of the
leading candidates for the realization of a practical qubit. Since the work of
Loss and DiVincenzo \cite{los98}, there have been a number of proposals
on how best to achieve the precise manipulations of such spins
required for the operation of quantum logic.  See, for example,
\cite{bur00,cal03,fen04,lov05}.

Whilst interest in electrostatic gating remains strong, the use of
lasers has several advantages in this role, most notably speed and
control.
Despite significant theoretical advances in this direction, there
has, as yet, been no experimental demonstration of
optically-controlled gating between electron spins in QDs.

In this paper, we describe an interaction mechanism to achieve just
this.  This qubit-qubit interaction is mediated by interdot
tunnelling of photo-excited carriers --- an area which has been the
subject of significant recent experimental advances
\cite{ort05,sti06}. Our results are of explicit relevance to the
current generation of vertically-stacked self-assembled InAs QDs,
but are also easily adaptable to the other dots, including
horizontally-coupled ones.

The interaction we describe has its origin in the so-called optical-RKKY effect, in which two electron spins are coupled via their
exchange interactions with optically-generated excitons in the
semiconductor bulk \cite{pie02}.  The coupling effect of these bulk
excitons between two electron spins in a double QD was examined in
Ref. \cite{ram05}.
Here we consider an interaction mediated, not by bulk excitons, but
by a single exciton confined in the same double QD structure.  We
describe a situation in which the excitonic electron is able to
tunnel between the dots and form delocalized `molecular' states. It
is the exchange interaction between this electron and the resident
qubit electrons that leads to an optically controlled gating.  This
gate, although not one of the standard quantum computation (QC)
gates, can be used to form a controlled-$Z$ operation when used
twice in conjunction with single qubit rotations, and is, in this way,
similar to the $\sqrt{\mathrm{SWAP}}$ gate.

The main factor limiting the speed with which operations can be
performed in this set-up comes from the `kinetic exchange' between
qubit spins which arises from the virtual tunneling of the qubit
electrons between dots. This is an important consideration in
vertically-stacked dots, and we show here that by choosing
appropriate dot parameters, we can make this effect small and still
obtain fast two-qubit operations.

Our proposal therefore offers an accessible path to the
demonstration of quantum logic in the solid state.  Furthermore, it
offers insight into the dynamics of interacting few-body systems in
confined nanostructures --- a topic of increasing experimental
relevance.

\begin{figure}[t]
  \begin{center}
    \epsfig{file=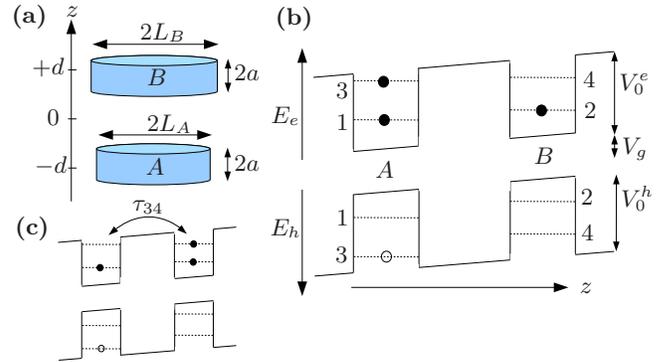,width=1\columnwidth}
    \caption{
    (colour online)
    {\bf (a)} Sketch of the two vertically-stacked quantum
    dots. Growth direction is the $z$ direction and the two dots are
    positioned symmetrically at $z=\pm d$.  The dots have heights $2a$
    and the width of the barrier between them is $b\equiv 2(d-a)$. The lower ($A$)
    and upper ($B$) dots have Darwin radii of $L_A$ and $L_B$
    respectively.
    {\bf (b)}  Two electronic and two hole levels in each dot take
    part in gate operation.  The lowest electron levels in each dot
    (labeled 1 and 2)
    are the two qubit levels and an electron permanently
    resides in each of them.  Laser illumination is tuned
    such that it creates an exciton in the excited levels of dot $A$ only
    (levels 3).
    {\bf (c)}  Due to a tunnel coupling $\tau_{34}$ between the dots
    and a resonance condition met through the tuning of the gate voltage $V_g$,
    the excitonic electron can
    tunnel back-and-forth.  The exchange interaction between this electron and the qubits
    gives rise to the optically-induced interaction between the qubits that can be used
    to perform a quantum gate.
      \label{f1}
    }
  \end{center}
\end{figure}

Figure \ref{f1} illustrates the basic principle at work here. The
two qubit electrons reside, one a piece, in the ground-state levels
of a vertical double-dot structure.  The dots have slightly
different diameters and a static gate voltage $V_g$ is applied in
the growth direction.  The distance between the dots is small enough
that tunneling can occur between the dots, but large enough to
suppress the direct exchange interaction between the two qubit
electrons. The tunneling of the qubit electrons into each others dot
is strongly suppressed by the energy required to doubly charge one
of the dots.

A circularly-polarized laser pulse is used to generate an exciton
into the lowest excited orbitals of dot A as shown in
Fig.~\ref{f1}b.  The gate voltage is tuned such that the energy of
this three-electron-and-one-hole configuration is resonant with that
shown in Fig.~\ref{f1}c, where the excited electron is in dot B.

The resonant tunneling of the electron between these configurations
creates a set of hybridised states in which the excitonic electron is
delocalized over both dots.  The qubit electrons experience a strong
intradot exchange interaction with this delocalized electron and
this mediates an effective interaction between the two qubits. By
controlling the laser parameters, this interaction can be harnessed
to perform quantum logic operations.

We begin our detailed treatment by describing the single-particle
states of the double dot.  This includes a calculation of the size
of the interdot tunneling elements.  For reasonable dot separations
we describe a situation where the qubits in the ground states are
effectively isolated from one another, but where the exciton
electron is free to tunnel.
We then detail the spectrum of the four-body system resulting from
the interplay of confinement, tunneling and the Coulomb interaction.
From this follows the quantum logic properties of the system.
Analytically, we derive an effective gate operator for the system
under adiabatic conditions.  We also use numerics to explore the
non-adiabatic capabilities. We show that gates equivalent to the
$\sqrt{\mathrm{SWAP}}$ gate can be obtained in times much less than
the exciton relaxation time and that the negative effects of
hole-mixing and spontaneous emission do not seriously affect these
results.

\section{Single-particle states}

The potential experienced by a conduction band (CB) electron in the
double-dot structure may be approximated as follows. In the growth
direction $z$ we have a double square-well potential with wells of
width $2a$ and depth $V_0$ centred at $\pm d$:
\beq
  V_z(z) =
  \left\{
  \begin{array}{cc}
  0 &  |z \pm d | \le a,\\
  V_0 & \mathrm{otherwise}.
  \end{array}
  \right.
\eeq
The width of the barrier between the dots (the parameter often
quoted in experiments) is given by $b\equiv 2(d-a)$.
Confinement in the $x$-$y$ plane is assumed harmonic with different
confinement energies above and below the $z=0$ plane corresponding
to dots $A$ and $B$:
\beq
  V_\rho({\bf r}) =
  \left\{
  \begin{array}{cc}
    1/2\, m_e \omega_A^2 \rho^2 &  z <0 , \\
   1/2\, m_e \omega_B^2 \rho^2 & z \ge 0.
  \end{array}
  \right.
\eeq
Here, $\rho = \sqrt{x^2 + y^2}$ is the radial coordinate, $m_e$ is
effective mass of the CB electron, and $\hbar \omega_X =
\hbar^2/(m_e L_X^2)$ is the confinement energy of dot $X=A,B$. The
Hamiltonian of the electron is then
\beq
  H_\mathrm{CB} = \frac{1}{2 m_e} {\bf p}^2 + V_z(z) + V_\rho({\bf r}) + E_g
  \label{H1}
  .
\eeq
where $E_g$ is the band-gap.

We will consider the two dots to be separated such that a
variational treatment in terms of wave functions localized in each
of the individual dots is appropriate.
Let us label with 1 and 3 the lowest two levels in dot $A$, and with
2 and 4 those in dot $B$.  Our starting  point, then, is the set of
four single-particle wave functions
\beq
  \phi _1({\bf r}) &=& \chi (z+d) \eta _s (\omega_A;\rho, \theta),
  \nonumber\\
  \phi _2({\bf r}) &=& \chi (z-d) \eta _s (\omega_B;\rho, \theta),
  \nonumber\\
  \phi _3({\bf r}) &=& \chi (z+d) \eta _p (\omega_A;\rho, \theta),
  \nonumber\\
  \phi _4({\bf r}) &=& \chi (z-d) \eta _p (\omega_B;\rho, \theta)
  \label{4phi}
  .
\eeq
In the growth direction, $\chi(z\mp d)$ are square-well
eigenfunctions centred at $z = \pm d$.   In the $xy$ plane, the
functions $\eta _s(\omega_X;\rho, \theta)$ and $\eta
_p(\omega_X;\rho, \theta)$ describe the ground and first-excited
Fock-Darwin wave functions for confinement energy $\omega_X$.  These
wave functions are given explicitly in Appendix \ref{Awfn}.
This set of wave functions is not orthogonal since, although the
overlaps $S_{12}=\ew{\phi_1|\phi_2}$ and $S_{34}=\ew{\phi_3|\phi_4}$
are small, they are finite.  We thus use their orthogonalized
counterparts $\psi_i(\mathbf{r})$ as outlined in the Appendix.

The evaluation of the Hamiltonian of Eq.~(\ref{H1}) in this
orthogonal basis gives us our approximate single CB-electron
Hamiltonian. Correct to first order in the overlaps $S_{12}\approx
S_{34}$, we have
\beq
  H_\mathrm{CB} =
  \rb{
    \begin{array}{cccc}
    E_1 & -\tau_{12} & 0 & 0 \\
    -\tau_{12} & E_2 +V_g & 0 & 0 \\
    0 & 0 & E_3 & -\tau_{34} \\
    0 & 0 & -\tau_{34} & E_4+V_g
    \end{array}
  }
  .
  \label{Hsp}
\eeq
The diagonal energies are given by
\beq
  E_1 &=& E_z + E_g + \hbar \omega_A;
  \quad  E_2 = E_z + E_g + \hbar \omega_B
  ;
  \nonumber\\
  E_3&=&E_z + E_g + 2\hbar \omega_A;
  \quad  E_4=E_z + E_g + 2\hbar \omega_B
  ,
\eeq
where $E_z$ is the energy resulting from the $z$ confinement and is
the same for both dots.

\begin{figure}[t]
  \begin{center}
    \epsfig{file=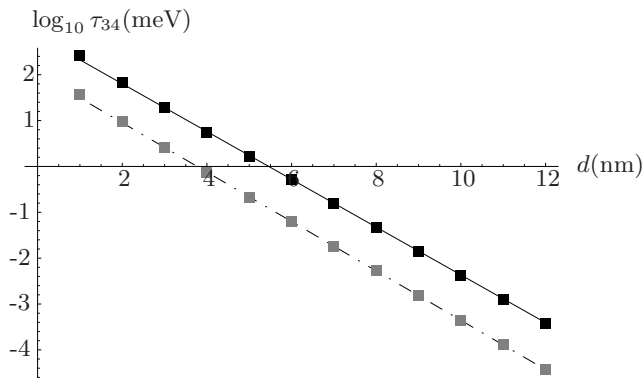,width=1\columnwidth}
    \caption{Logarithm of $\tau_{34}$, the tunneling amplitude between
    excited states in different dots as a function of the interdot
    distance $d$.
    The solid black line shows the approximate result of Eq.~(\ref{eqt}) and
    the black points show a numerical evaluations of full
    single-particle matrix element. The dashed
    line and grey boxes show the analogous quantities for heavy-hole
    tunneling.
    The dot height is $2a = 2$~nm and the radial confinement energies
    were $\hbar \omega_A^e = 35$~meV and $\hbar \omega_B^e = 30$~meV
    for all carriers.  The vertical confinement depth was
    $V_0^e = 680$~meV for electrons and $V_0^h = 100$~meV for holes.
    We took the electron mass to be 0.04$m_0$ in InAs and 0.067$m_0$
    in the surrounding GaAs matrix.  Heavy holes were masses $m_{hz} = 0.34m_0$
    and $m_{h\perp} =  0.04m_0$ for both InAs and GaAs.
      \label{figt34}
    }
  \end{center}
\end{figure}

The quantities $\tau_{12}$ and $\tau_{34}$ are ground- and
excited-state tunneling amplitudes. They may be approximated as
\beq
  \tau_{12} \approx  \tau_{34} \approx
  4 a  V_0 B C\frac{ L_A L_B}{L_A^2 + L_B^2} e^{-2 \beta d}
  \label{eqt}
  ,
\eeq
where $\beta$, $B$, and $C$ are functions of $V_0$ determined by the continuity of the $z$-direction wave function (see appendix).
The dependence of $\tau_{34}$ on the interdot distance is
illustrated in Fig.~\ref{figt34} for typical parameters.
Current experiments have largely operated at small interdot spacings
where the tunneling is large.  For example, Ortner and coworkers
\cite{ort05} measured a value of $\tau_{12}\approx 13$~meV for
$d\approx 3.5$~nm, which is in good agreement with the results
plotted in Fig.~\ref{figt34}.
For our purposes, we require a much smaller tunneling such that
localized states are still well defined. We will consider
separations in the range of  $d\approx 8$~nm ($b\approx 14$~nm),
where the tunneling amplitude lies in the range 10-100~$\mu$eV.

In Eq.~(\ref{Hsp}) we have also taken into account the effects of
the gate voltage applied in the $z$ direction by incorporating a
first-order shift in the energy levels of dot $B$ by an amount $V_g$
\cite{sti06}.


In InAs dots, the valence-band states have a predominantly
heavy-hole (HH) character.
The single-particle Hamiltonian for such holes follows in exactly
the same way as above except that the parameters differ.  In
particular, the HH mass in the growth direction, $m_{hz}$, is
different to that in the $xy$ plane $m_{h\perp}$.
The heavier mass of the HH in the $z$ direction, as compared with
the electron, means that the HHs are more localized than the
electrons. This in turn means that the HH tunneling amplitudes
between the dots are approximately one order of magnitude smaller
than those for the electrons, as illustrated in Fig. \ref{figt34}.

\section{Light-Matter Interaction}

The multi-particle basis states for our problem are chosen by filling the appropriate single-particle levels.  Prior to the application of the laser, there are two electrons in the system and these reside in the ground-state levels of the dots.  Interaction with the laser generates an additional electron-hole pair in the excited-levels of the double-dot system.

We consider here illumination with a $\svec_+$ circularly polarized laser propagating
in the growth direction.  Conservation of angular momentum means that this excites an exciton consisting of a spin-down electron and a spin-up heavy hole.  Here, we have neglected hole-mixing and assumed that the in-plane magnetic field is zero --- points which we return to later.  The laser is tuned such that it creates the exciton in the excited levels of the dots and not in the ground-state levels where the qubit electrons reside.  In principle, illumination can create an exciton with a hole in either dot

The block structure of the Hamiltonian for the interacting system 
reads
\beq
  H_{\mathrm{total}} = 
  \rb{
  \begin{array}{ccc|cc}
    X^{(D)} & T^e_{X4} & T^e_{X3} & 0 & \Omega^{(D)} \\
    {T^e_{X4}}^\dag & X^{(4)} & T^h & \Omega^{(4)} & 0 \\
    {T^e_{X3}}^\dag & {T^h}^\dag & X^{(3)} & \Omega^{(3)} & 0 \\
    \hline
    0 & {\Omega^{(4)}}^\dag & {\Omega^{(3)}}^\dag & G & T_G^e \\
    {\Omega^{(D)}}^\dag & 0 & 0 & {T^e_G}^\dag & D
  \end{array}
   }
   \label{Htot}
   .
\eeq
The two blocks in the bottom righthand corner, $G$ and $D$, describe states consisting of the two resident electrons only --- in $G$ the electrons are separated one in each dot; in $D$, one of the dots is doubly occupied.
These two blocks are coupled via the electron-tunneling terms of block $T^e_G$.
The blocks $X^{(3)}$ and $X^{(4)}$ describe states with two resident electrons plus the exciton, with the superscript referring to the dot-level in which the hole is located.  These two blocks are coupled by the hole-tunneling terms of block $T^h$.
Block $X^{D}$ describes states where an exciton is present and one of the ground-state levels is doubly occupied.  This block is coupled to the other exciton-containing states through the electron tunneling blocks $T^e_{X3,4}$.
Finally, the blocks $\Omega^{(i)}$; $i=3,4,D$ contain the light-matter interaction terms, and provide the connexion between the ground and excited sectors.
It should be noted that the effects of Coulomb interaction are incorporated into the diagonal blocks of $H_\mathrm{total}$, and that the exciton blocks, $X^{(i)}$  also contain single-electron tunneling terms.

We can make several approximation that reduce the complexity of $H_\mathrm{total}$ significantly.
Firstly we assume that, whereas in principle the laser can excite an electron hole-pair in either of the dots, we assume that the two dots are sufficiently distinct in size that, in fact they are only generated in dot $A$.  We will later chose the difference in confinement energies of the two dots to be $\approx 5$~meV, so this is certainly valid.
We can also neglect the generation of the indirect excitons such as $e^\dag_{3 \downarrow} h^\dag_{4 \Uparrow}$, as these have very small transition matrix elements.
This means that the light-matter interaction is well approximated as
\beq
  H_{\mathrm{int}} = \Omega(t) e^{-i \omega t}
  e^\dag_{3 \downarrow} h^\dag_{3 \Uparrow} + \mathrm{H.c.}
  ,
  \label{LMWW}
\eeq
where $\Omega(t)$ is the time-dependent Rabi frequency of the exciton transition.  
The block $\Omega^{(4)}$, corresponding to the generation of excitons in dot $B$, is therefore zero.  Furthermore, hole-tunneling is very effectively suppressed in this system --- not only are the  hole-tunnelling matrix elements much smaller than their electronic counterparts (see previous section), but under the operating conditions of our device, to be described below, these transitions are far off-resonant.  We therefore treat the block $X^{(4)}$, with a hole in dot $B$, as being effectively decoupled from the dynamics that we are interested in.

Finally, transitions to states in which the dot levels are doubly occupied are suppressed by the enhanced Coulomb interaction of such states.  That this is still the case even with different dot sizes and in the presence of applied field is jusitified in the next section.  This means that the blocks $D$ and $X^D$ are decoupled from the rest of the system over the relevant timescales.
Taken together, these approximations mean that the system can be described with just the portion of the Hamiltonian $H_\mathrm{total}$ consisting of the blocks $G$, $X^{(3)}$ and $\Omega^{(3)}$.  Since these blocks have significant internal structure themselves, the solution of this problem is still by no means trivial.

\section{Ground-state sector: Isolated Qubits}

We continue our treatment by temporarily reinstating block $D$ and consider all states in which there are only two electrons in the system.  This is the ``ground-state sector'', and is described by blocks $G$, $D$ and $T_G^e$ above.  It is clear that only states with one electron in each dot represent good two-qubit states and that double-occupancy of either dot, as in block $D$, is a source of error.  It is therefore the purpose of this section to describe this sector and to show the extent to which block $D$ can be neglected and the qubits thought of as isolated.

We can decompose the proper qubit sector as $G=\mathrm{Diag}\left\{G_{-1},G_0,G_{+1}\right\}$, where the subblocks are labelled by the $z$-projection of the total electron spin.  Block $G_{-1}$ corresponds to the energy of the two-qubit state $e^\dag_{1\downarrow}e^\dag_{2\downarrow}\ket{0}$,
$G_{+1}$ corresponds to $e^\dag_{1\uparrow}e^\dag_{2\uparrow}\ket{0}$, and $G_0$ consists of the two qubit states with the electrons having opposite spins: $e^\dag_{1 \downarrow} e^\dag_{2 \uparrow}\ket{0}$ and $ e^\dag_{1
\uparrow} e^\dag_{2 \downarrow}\ket{0}$.

The exclusion principle determines that there are only two states in block $D$, namely $e^\dag_{1 \downarrow}
e^\dag_{1 \uparrow}\ket{0} $ and $e^\dag_{2 \downarrow} e^\dag_{2
\uparrow}\ket{0} $.  Conservation of electron spin angular momentum means that only block $G_0$ couples to the block $D$.  We can therefore consider the effects of double occupation  by concentrating on the following Hamiltonian
\beq
  H_{GD}=
  \rb{
  \begin{array}{cc}
     G_{0} & T^e_G \\ 
     T^e_G & D 
  \end{array}
   }
   ,
\eeq
each block of which is a two-by-two-matrix.

To assess the effects of double occupancy it is necessary to consider the Coulomb interaction between the confined carriers, in this case electrons.  The electron-electron matrix elements between single-particle orbitals are given by
\beq
  V^{eeee}_{ijkl} = E_C \int d^3 {\bf r_1} d^3 {\bf r}_2
    \frac{
      \psi_i^e({\bf r}_1)
      \psi_j^e({\bf r}_2)
      \psi_k^e({\bf r}_2)
      \psi_l^e({\bf r}_1)
    }
    {|{\bf r}_1-{\bf r}_2|}
    ,
\eeq
with $E_C= e^2/(4\pi \epsilon_0 \epsilon_R)$.  We define similar
quantities  $V^{ehhe}_{ijkl}$ for electron-hole interactions and will
only consider electron-hole direct interaction as $e$-$h$ exchange
effects are negligible.

Including both Coulomb interactions and single-particle terms, the
Hamiltonian of the ground-state sector in question is 
\begin{widetext}
\beq
  H_{GD} =
  \rb{
    \begin{array}{cccc}
    E^G +V_g& J_{12} & -\tau_{12}^e & -\tau_{12}^e\\
    J_{12} & E^G +V_g & \tau_{12}^e & \tau_{12}^e\\
    -\tau_{12}^e & \tau_{12}^e
      & E^G-V_{1221}^{eeee} +V_{1111}^{eeee} - \Delta E & 0\\
    -\tau_{12}^e& \tau_{12}^e & 0
      & E^G-V_{1221}^{eeee} +V_{2222}^{eeee} + \Delta E +2V_g
    \end{array}
  }
  ,
\eeq
\end{widetext}
where we have defined $E^G = E_1^e + E_2^e + V_{1221}^{eeee}$,
$\Delta E = E_2^e-E_1^e$, and $J_{12}=V^{eeee}_{1212}$ is the
magnitude of ground-state exchange interaction.
This exchange term can be approximated as
\beq
  J_{12} \approx
  \frac{8 E_C \sqrt{\pi} C^4}
    {L_A^2 L_B^2 \rb{L_A^{-2}+L_B^{-2}}^{3/2}}
  (d-a)^2 e^{-4 \beta d}
  \label{eqJ12}
  .
\eeq
For the dot separations in which we are interested here ($d \approx
8$~nm), this exchange energy is of the order $10^{-5}$~meV and can
thus be neglected.


The tunneling amplitude $\tau_{12}$ means that double-occupancy of
the qubit levels can, in principle, occur, and  this would clearly
be deleterious.  However, if the doubly occupied states are much
higher in energy than the singly-occupied ones, double occupancy
will be suppressed . This requires that the energy differences
satisfy
\beq
  Q_1 \equiv
    V_{1111}^{eeee}-V_{1221}^{eeee} - \Delta E - V_g
  \gg \tau_{12}^e
  ,
  \nonumber
  \\
  Q_2 \equiv
    V_{2222}^{eeee}-V_{1221}^{eeee} + \Delta E + V_g
  \gg \tau_{12}^e
  .
  \label{Q1Q2}
\eeq
With these inequalities satisfied, the effect of double occupancy is
reduced to imparting a small residual `kinetic exchange' interaction
on the fixed qubits. To second order in $\tau_{12}$ this kinetic
exchange energy may be approximated as $\tau_{12}^2/Q$, with
$Q^{-1}= Q_1^{-1}+Q_2^{-1}$. The important time-scale,
\beq
  T_Q = \hbar Q / \tau_{12}^2
  ,
\eeq
follows accordingly such that,  for times $t\ll T_Q$, the qubits are
effectively isolated from one another.

We will see later that device operation requires that the offset
$V_g$ should be set as
\beq
  V_g
  &=& E^e_3 - E^e_4
  +V^{eeee}_{1331}-V^{eeee}_{2442}
  +V^{eeee}_{2332} - V^{eeee}_{1441}
  \nonumber\\
  &&
  +V^{ehhe}_{4334} - V^{ehhe}_{3333}
  +J_{24}-J_{13}
  .
\eeq
As the Coulomb elements are approximately symmetric between the
dots, we can approximate $Q_i$ as
\beq
  Q_1 &\approx&
   V_{1111}^{eeee}-V_{1221}^{eeee}
   + V_{3333}^{ehhe} - V_{4334}^{ehhe} + \Delta E
   ,
  \nonumber\\
  Q_2 &\approx&
    V_{2222}^{eeee}-V_{1221}^{eeee}
   - V_{3333}^{ehhe} + V_{4334}^{ehhe} - \Delta E
   .
\eeq
In order to obtain as long a time $T_Q$ as possible, we need to
maximize $Q$, and this is achieved when $Q_1=Q_2$.  This in turn
means that the optimal difference between the  confinement energies
of the two dots should be $
  \Delta E \approx - V_{3333}^{ehhe} + V_{4334}^{ehhe},
$ in which case $Q = Q_1/2 \approx
1/2(V_{1111}^{eeee}-V_{1221}^{eeee})$.
Typical parameters give $\Delta E\approx Q \approx 5$~meV and thus
$T_Q=1.3$~ns for a tunneling amplitude of $\tau_{12} = 0.05$~meV,
and $T_Q=82$~ps for $\tau_{12} = 0.2$~meV.
Providing then that the operation of our gate is much faster than $T_Q$, we can neglect the effects of double occupancy in the ground-state sector.  A similar analysis shows that tunneling of ground-state electrons is suppressed to a similar
degree also in the presence of the exciton.  This means that, with the above proviso, we are justified in neglecting blocks $D$ and $X^D$ from $H_\mathrm{total}$.

The other important time-scale for this system is the relaxation rate
of the exciton.  Experiments have shown that this quantity is of the
order of $T_\mathrm{rel}\approx 1$~ns or greater \cite{cor02,war05}.  This, therefore, is of the same
order of magnitude as $T_Q$ for $\tau_{12}= 0.05$~meV.

\section{Exciton states}

We now consider states with the exciton, and since the hole is always localized in state 3, we consider only block $X^{(3)}$.  We can decompose this block into subsectors based on the $z$-projection of the total electron spin: $X^{(3)} = \mathrm{Diag}\left\{X^{(3)}_{3/2},X^{(3)}_{1/2},X^{(3)}_{-1/2},X^{(3)}_{-3/2}\right\}$.  Since there are no spin-flip processes effective over the time scale of a typical
gate operation, these sectors are independent and each excitonic sector couples exclusively to one of sub-blocks of $G$.  This means that the total Hamiltonian for the system can be written
\begin{widetext}
\beq
  H_{\mathrm{total}} = 
  \rb{
  \begin{array}{cc|cc|cc|c}
    X^{(3)}_{-3/2} & \Omega^{(3)}_{-3/2} & 0  & 0 & 0 & 0 & 0 \\
    {\Omega^{(3)}_{-3/2}}^\dag & G_{-1} & 0 & 0 & 0 & 0 & 0 \\
    \hline
    0 & 0 & X^{(3)}_{-1/2} & \Omega^{(3)}_{-1/2} & 0 & 0 & 0 \\
    0 & 0 & {\Omega^{(3)}_{-1/2}}^\dag & G_{0} & 0 & 0 & 0 \\
    \hline
    0 & 0 & 0 & 0 & X^{(3)}_{1/2} & \Omega^{(3)}_{1/2} & 0 \\
    0 & 0 & 0 & 0 & {\Omega^{(3)}_{1/2}}^\dag & G_{+1} & 0\\
    \hline
     0 & 0 & 0 & 0 & 0 & 0 & X^{(3)}_{+3/2}
  \end{array}
   }
   \label{Hblock}
\eeq
\end{widetext}
Let us label each of these interacting subblocks with the electron-spin projection of the exciton state.  We therefore  have  $H_\mathrm{total} = \mathrm{Diag}\left\{H^{(-3/2)},H^{(-1/2)},H^{(1/2)},H^{(3/2)}\right\}$.  Block $H^{(3/2)}$ is clearly decoupled from the dynamics of the system and can be ignored.

Let us consider the sector described by $H^{(-3/2)}$, in which both qubits are spin-down and
thus parallel to the electronic spin of the exciton.  Only three
states then make up this sector: the photo-excited state
$\ket{\mathrm{I}} =
    e^\dag_{1 \downarrow} e^\dag_{2  \downarrow}
    e^\dag_{3  \downarrow} h^\dag_{3 \Uparrow} \ket{0}$, the same configuration but with the electron tunnelled
$\ket{\mathrm{II}} =
    e^\dag_{1 \downarrow} e^\dag_{2  \downarrow}
    e^\dag_{4  \downarrow} h^\dag_{3 \Uparrow} \ket{0}$, and the qubit ground-state
$\ket{\mathrm{III}} =
    e^\dag_{1 \downarrow} e^\dag_{2  \downarrow}
    \ket{0}
$.

The diagonal elements of the Hamiltonian for the first two states
are
\beq
  H^{(-3/2)}_{\mathrm{I},\mathrm{I}}&=& C_{\mathrm{I}}+ V_g -
  J_{13},
  \nonumber\\
  H^{(-3/2)}_{\mathrm{II},\mathrm{II}} &=& C_{\mathrm{II}} + 2 V_g-
  J_{24},
\eeq
where $C_{\mathrm{I}}$ is the sum of all single-particle energies
and direct Coulomb interactions for state $\ket{I}$:
\beq
  C_{\mathrm{I}} &=&
  E^e_1 + E^e_2 + E^e_3 + E^h_3
  \nonumber\\
  &&
  + V^{eeee}_{1221} + V^{eeee}_{1331} + V^{eeee}_{2332}
    \nonumber\\
  &&
  - V^{ehhe}_{1331} - V^{ehhe}_{2332} - V^{ehhe}_{3333}
  ,
\eeq
and similarly for $C_{\mathrm{II}}$.  The energies $J_{13}=V^{eeee}_{1313}$ and $J_{24}=V^{eeee}_{2424}$ are the intradot exchange interaction strengths, and all other
exchange interactions are negligible.

In order that the excess
electron can tunnel between the two configurations I and II, we set
the voltage $V_g$ across the device such that we align
$H^{\downarrow \downarrow}_{\mathrm{I},\mathrm{I}} = H^{\downarrow
\downarrow}_{\mathrm{II},\mathrm{II}} \equiv E^X$.  This requires
\beq
  V_g &=& C_{\mathrm{I}}-C_{\mathrm{II}} - J_{13} + J_{24}
  \nonumber\\
  &=& E^e_3 - E^e_4
  +V^{eeee}_{1331}-V^{eeee}_{2442}
  +V^{eeee}_{2332} - V^{eeee}_{1441}
  \nonumber\\
  &&
  +V^{ehhe}_{4334} - V^{ehhe}_{3333}
  +J_{24}-J_{13}
  ,
  \label{V_g}
\eeq
which yields $V_g\approx 5$~meV for our parameters.

With this condition, the Hamiltonian for this sector reads
\beq
H^{(-3/2)} =
  \rb{
    \begin{array}{ccc}
    E^X& - \tau_{34}
        & \Omega e^{-i\omega t}\\
    -\tau_{34} & E^X & 0 \\
    \Omega e^{i \omega t} & 0 & E^G
    \end{array}
  }
  .
\eeq
Moving to a basis of tunneling eigenstates for the exciton levels,
$2^{-1/2}(\ket{\mathrm{I}} \pm \ket{\mathrm{II}})$, we have
\beq
H^{(-3/2)} =
  \rb{
    \begin{array}{ccc}
    E^X - \tau_{34} & 0
        & \Omega e^{-i\omega t}/\sqrt{2}\\
    0 & E^X+ \tau_{34}& \Omega e^{-i\omega t}/\sqrt{2} \\
    \Omega e^{i \omega t}/\sqrt{2} & \Omega e^{i\omega t}/\sqrt{2}
      & E^G
    \end{array}
  }
  .
\eeq
We tune the laser frequency such that, apart from a small detuning
$\delta$, it is on resonance with the transition from the ground
state and the lowest tunnel coupled exciton state with energy $E_X-
\tau_{34}$:
\beq
  \omega + \delta =  E^X-\tau_{34}-E^G.
\eeq
Finally, we move to a rotating frame 
\beq 
  H^{(m)} \rightarrow H^{(m)}_R = RH^{(m)}R^\dag + i \dot{R}R^\dag
  ,
\eeq 
with, in this case, $R=\mathrm{Diag}\left\{e^{i(\omega+E_G)t},e^{i(\omega+E_G)t},e^{iE_Gt}\right\}$
to obtain
\beq
  H^{(-3/2)}_R &=&
  \rb{
    \begin{array}{ccc}
    \delta & 0 & \Omega /\sqrt{2}\\
    0 & \delta + 2\tau_{34}& \Omega /\sqrt{2} \\
    \Omega^*/\sqrt{2} & \Omega^* /\sqrt{2}
      & 0
    \end{array}
  }
 \label{HtDD}
 .
\eeq

We now consider the Hamiltonian, $H^{(-1/2)}$, for the sector in which the qubit spins have opposite
directions.
This is the important sector as it is here that the exchange between
exciton and qubit electrons is manifest.  Taking into account both
electron tunneling and electron exchange, there are six exciton
states in this sector.  With intra-dot exchanges $J_{13}$ and
$J_{24}$ diagonal, these are 
\beq
  \ket{\mathrm{I}}&=&
    e^\dag_{1 \downarrow} e^\dag_{2  \uparrow}
    e^\dag_{3  \downarrow} h^\dag_{3 \Uparrow} \ket{0}
    ,
  \nonumber\\
  \ket{\mathrm{II}} &=& 2^{-1/2}
    \rb{
      e^\dag_{1 \uparrow} e^\dag_{2  \downarrow}
      e^\dag_{3  \downarrow}
      +
      e^\dag_{1 \downarrow} e^\dag_{2  \downarrow}
      e^\dag_{3  \uparrow}
    }h^\dag_{3 \Uparrow} \ket{0}
    ,
  \nonumber\\
  \ket{\mathrm{III}} &=&
    e^\dag_{1 \uparrow} e^\dag_{2  \downarrow}
    e^\dag_{4  \downarrow} h^\dag_{3 \Uparrow} \ket{0}
    ,
    \nonumber\\
  \ket{\mathrm{IV}}&=& 2^{-1/2}
    \rb{
      e^\dag_{1 \downarrow} e^\dag_{2  \uparrow}
      e^\dag_{4  \downarrow}
      +
      e^\dag_{1 \downarrow} e^\dag_{2  \downarrow}
      e^\dag_{4  \uparrow}
    }h^\dag_{3 \Uparrow} \ket{0}
    ,
  \nonumber\\
  \ket{\mathrm{V}}&=& 2^{-1/2}
    \rb{
      e^\dag_{1 \uparrow} e^\dag_{2  \downarrow}
      e^\dag_{3  \downarrow}
      -
      e^\dag_{1 \downarrow} e^\dag_{2  \downarrow}
      e^\dag_{3  \uparrow}
    }h^\dag_{3 \Uparrow} \ket{0}
    ,
  \nonumber\\
   \ket{\mathrm{VI}}&=& 2^{-1/2}
    \rb{
      e^\dag_{1 \downarrow} e^\dag_{2  \uparrow}
      e^\dag_{4  \downarrow}
      -
      e^\dag_{1 \downarrow} e^\dag_{2  \downarrow}
      e^\dag_{4  \uparrow}
    }h^\dag_{3 \Uparrow} \ket{0}.
\eeq
The first four states are degenerate under the intradot exchanges,
and states 5 and 6 are split from the rest by an energy of
$2J_{13}\approx 2J_{24}\approx 6.5$~meV.  This means that they are
effectively decoupled and can be neglected henceforth.

We can then proceed to the basis of tunneling eigenbasis defined by
\beq
 \ket{\mathrm{I}'} &=&
     1/\sqrt{6} \ket{\mathrm{I}}
   + 1/\sqrt{3} \ket{\mathrm{II}}
   + 1/\sqrt{6} \ket{\mathrm{III}}
   + 1/\sqrt{3} \ket{\mathrm{IV}}
   ,
 \nonumber\\
 \ket{\mathrm{II}'} &=&
     1/\sqrt{3} \ket{\mathrm{I}}
   - 1/\sqrt{6} \ket{\mathrm{II}}
   - 1/\sqrt{3} \ket{\mathrm{III}}
   + 1/\sqrt{6} \ket{\mathrm{IV}}
   ,
 \nonumber\\
 \ket{\mathrm{III}'} &=&
   - 1/\sqrt{3} \ket{\mathrm{I}}
   + 1/\sqrt{6} \ket{\mathrm{II}}
   - 1/\sqrt{3} \ket{\mathrm{III}}
   + 1/\sqrt{6} \ket{\mathrm{IV}}
   ,
 \nonumber\\
 \ket{\mathrm{IV}'} &=&
   - 1/\sqrt{6} \ket{\mathrm{I}}
   - 1/\sqrt{3} \ket{\mathrm{II}}
   + 1/\sqrt{6} \ket{\mathrm{III}}
   + 1/\sqrt{3} \ket{\mathrm{IV}}
   .
 \nonumber\\
\eeq
In the basis of these four states plus the two ground states
$e_{1\downarrow}^\dag e_{2 \uparrow}^\dag$ and $e_{1\uparrow}^\dag
e_{2 \downarrow}^\dag$ the Hamiltonian in the rotating frame reads
\begin{widetext}
\beq
  H^{(-1/2)}_R &=&
  \rb{
  \begin{array}{cccccccc}
   \delta & 0 & 0 & 0
      & \Omega(t) /\sqrt{6}
      & \Omega(t) /\sqrt{6}\\
    0 & \delta + \tau^e_{34}/2 & 0 & 0
      & \Omega(t) /\sqrt{3}
      & -\Omega(t) /\sqrt{12}\\
    0 & 0 & \delta  +3\tau^e_{34}/2 & 0
      & -\Omega(t) /\sqrt{3}
      & \Omega(t) /\sqrt{12}\\
    0 & 0 & 0 & \delta + 2\tau^e_{34}
      & -\Omega(t) /\sqrt{6}
      & -\Omega(t) /\sqrt{6} \\
   \Omega^*(t) /\sqrt{6}&
       \Omega^*(t) /\sqrt{3}&
     - \Omega^*(t) /\sqrt{3} &
     - \Omega^*(t) /\sqrt{6} &
     0 & 0\\
   \Omega^*(t) /\sqrt{6} &
     - \Omega^*(t) /\sqrt{12} &
       \Omega^*(t) /\sqrt{12} &
     - \Omega^*(t) /\sqrt{6} &
     0 & 0\\
  \end{array}
  }
  \label{HtUD}
  ,
\eeq
\end{widetext}
where the rotating frame is defined with frequency $\omega+E_G$ for all excited states and $E_G$ for the two ground states.

Analysis of the final sector, in which both qubit spins point up,
proceeds exactly as for $H^{(-1/2)}$ above, except with all
electron spins flipped and only a single ground state
$e_{1\uparrow}^\dag e_{2\uparrow}^\dag \ket{0}$. The resulting
Hamiltonian, $H^{(1/2)}$, is the same as that of Eq.~(\ref{HtUD}) but with the fifth state omitted.

Taking these results together, and ignoring for the moment the light-matter coupling, we see that the excitonic sector consists of four energy levels, of which the inside and outside pairs are doubly and triply degenerate, respectively.  In each of
these states the excitonic electron is delocalized over the entire
structure, and is thus capable of mediating an exchange interaction between the qubits. Note that the optical coupling between these states and the qubit
ground states shows a variety of different coupling strengths, as
determined by the forefactors of $\Omega$ in the preceding
Hamiltonians.

\section{Quantum Gates}

Having derived these Hamiltonians, we now show how the interactions
they entail may be used to perform two-qubit operations.  We begin
by considering the system in the adiabatic limit in which the
exciton levels are but virtually populated.  This approach yields an
effective gate operator $U_\mathrm{eff}$ acting only on the qubit
space.
Due to the complexity of the exciton structure, $U_\mathrm{eff}$
does not have the form of one of the standard QC gates.  However, we
give the relationship between $U_\mathrm{eff}$ and the
controlled-phase (CPHASE) gate, which makes clear the QC capacity of
$U_\mathrm{eff}$.
We then consider non-adiabatic operation of the system and show how
this can improve operation times.


Second-order Rayleigh-Schr\"{o}dinger perturbation theory of Raman
processes can be used to derive an effective Hamiltonian for the
qubit sector through the elimination of the exciton levels.
In the basis $\left\{\ket{\downarrow \downarrow},\ket{\downarrow
\uparrow},\ket{\uparrow \downarrow},\ket{\uparrow \uparrow}\right\}$
we obtain the effective Hamiltonian
\beq
 {\cal H}_\mathrm{eff} &=&
 -\frac{|\Omega(t)|^2}{\delta}
 \rb{
   \begin{array}{cccc}
     x_1 & 0 & 0 & 0 \\
     0 & x_2 & y & 0 \\
     0 & y   & x_3 & 0 \\
     0 & 0 & 0 & x_3
   \end{array}
 }
 = -\frac{|\Omega(t)|^2}{\delta} \hat{M}
 \label{Heff}
 .
\eeq
with elements
\beq
  x_1 &=&
  \frac{\delta}{2}
  \left\{
    \frac{1}{\delta}
    + \frac{1}{\delta+2\tau^e_{34}}
  \right\}
  ,
  \nonumber\\
    x_2 &=&
  \frac{\delta}{6}
  \left\{
    \frac{1}{\delta}
    + \frac{2}{\delta+\tau^e_{34}/2}
  \right.
  \nonumber\\
  &&~~~~~~~~~~~~~
  \left.
    + \frac{2}{\delta+3\tau^e_{34}/2}
    + \frac{1}{\delta+2\tau^e_{34}}
  \right\}
  ,
  \nonumber\\
  x_3 &=&
  \frac{\delta}{12}
  \left\{
    \frac{2}{\delta}
    + \frac{1}{\delta+\tau^e_{34}/2}
  \right.
  \nonumber\\
  &&~~~~~~~~~~~~~
  \left.
    + \frac{1}{\delta+3\tau^e_{34}/2}
    + \frac{2}{\delta+2\tau^e_{34}}
  \right\}
  ,
  \nonumber\\
  y &=&
  \frac{\delta}{6}
  \left\{
    \frac{1}{\delta}
    - \frac{1}{\delta+\tau^e_{34}/2}
  \right.
  \nonumber\\
  &&~~~~~~~~~~~~~
  \left.
    - \frac{1}{\delta+3\tau^e_{34}/2}
    + \frac{1}{\delta+2\tau^e_{34}}
  \right\}
  .
  \label{xy}
\eeq
The validity of this Hamiltonian is conditioned on the standard
adiabatic conditions of $ T\delta / \hbar \gg 1$, with $T$ the
operation duration, to avoid populating the trion levels, and
$|\Omega(t)| /\delta \ll 1$ such that perturbation theory is valid.
Under these conditions, the time-evolution operator of the qubit
sector may be approximated as
$
  U_\mathrm{eff}(t) \approx e^{- i \Lambda \hat{M}}
$ with $\Lambda = - \int dt |\Omega(t)|^2 / \hbar \delta$. We will use a
Gaussian pulse shape, $
  \Omega(t) = A \exp\rb{-t^2/2T^2}
$, such that $\Lambda = -A^2 \sqrt{\pi} T/\hbar \delta$.
Evaluating the matrix exponential we find that $U_\mathrm{eff}$ can
be written as
\begin{widetext}
\beq
  U_\mathrm{eff} =
   \rb{
     \begin{array}{cccc}
     e^{i \kappa_1} & 0 & 0 & 0 \\
     0 & e^{i q_-} \cos\Theta & -i e^{i (q_-+q_+)/2} \sin\Theta & 0 \\
     0 &  -i e^{i (q_-+q_+)/2} \sin\Theta&  e^{i q_+} \cos\Theta& 0 \\
     0 & 0 & 0 & e^{i \kappa_3}
   \end{array}
    }
  \label{Ueff}
  .
\eeq
\end{widetext}
where $\kappa_1 = -\Lambda x_1$, $\kappa_3 = -\Lambda x_3$,
$\sin\Theta=\sin\phi \cos\theta$ with $\phi =
\Lambda/2\sqrt{(x_2-x_3)^2 + 4y^2}$ and $\tan\theta=(x_2-x_3)/2y$,
and where $q_\pm=\pm\chi - \Lambda(x_2+x_3)/2$ with
$\tan\chi=\sin\theta \tan \phi$.

As is evident, $U_\mathrm{eff}$ is not a standard QC gate.  However,
we now make the connection between $U_\mathrm{eff}$ and the familiar
CPHASE gate
\beq
  C(\Phi) = \text{Diag}\rb{e^{i\Phi}, 1, 1, 1}
  ,
\eeq
where $\Phi$ is the controlled phase.  Let us define the
single-qubit operation $S_i(\alpha) = \text{Diag}\rb{e^{i\alpha},
1}$ acting on qubit $i$, and
the generalized SWAP gate
\beq
  W(\Theta) =
  \left(
  \begin{array}{cccc}
    1 &  0  & 0  &  0 \\
    0   & \cos\Theta  &   -i\sin\Theta  &  0 \\
    0  &  -i\sin\Theta   &  \cos\Theta  &   0 \\
    0 & 0  &  0  & 1 \\
  \end{array}
  \right)
  .
\eeq
Setting $\Theta=\pi/2$ in $W(\Theta)$ yields a SWAP gate, and
setting $\Theta=\pi/4$ gives a $\sqrt{\mathrm{SWAP}}$ entangling
gate.
In terms of these operators the physical gate $U_\mathrm{eff}$ can
be written as
\beq
U_\text{eff} &=&
  e^{i\kappa_3}
  S_1(\frac{q_--\kappa_3}{2})
  S_2(\frac{q_+-\kappa_3}{2})
  \nonumber\\
  &&~~~~
  \times
  C(\Phi/2)W(\Theta)
  \nonumber\\
  &&~~~~~~~~
  \times
  S_1(\frac{q_--\kappa_3}{2})
  S_2(\frac{q_+-\kappa_3}{2}).
\eeq
where
\beq
  \Phi &=& 2\rb{\kappa_1 + \kappa_3 - q_- - q_+}
  \nonumber\\
  &=& -4 \Lambda y
  \nonumber\\
  &=&
  \frac{2}{3}
  \frac{A^2 \sqrt{\pi}T}{\hbar}
  \left\{
    \frac{1}{\delta}
    - \frac{1}{\delta+\tau^e_{34}/2}
  \right.
  \nonumber\\
  &&~~~~~~~~~~~~~
  \left.
    - \frac{1}{\delta+3\tau^e_{34}/2}
    + \frac{1}{\delta+2\tau^e_{34}}
  \right\}
  \label{Phi}
  .
\eeq
The gate $U_\mathrm{eff}$ therefore contains a product of two
entangling operations, $C(\Phi/2)$ and $W(\Theta)$.

We can isolate the CPHASE gate from the product $CW$ by observing
that the key unitary transformation of $W$ by the single-qubit
operation $S'_2(\pi) =\text{Diag}\rb{1,e^{-i\pi}}$ gives $S'_2(\pi)
W {S'_2}^\dag(\pi) = W^\dagger$. Since $S_2'$ commutes with $C$, the
concatenation of $CW$ and its transform yields,
\beq
  C(\Phi/2)W(\Theta) S'_2(\pi) C(\Phi/2)W(\Theta) S'_2(\pi) = C(\Phi).
  \label{CWCW}
\eeq
Note that the transformation is independent of $\Theta$ and  can be
carried out without the knowledge of its value.
From this it follows that the composite operation of two
applications of $U_\mathrm{eff}$ interspersed with appropriate
single-qubit operations is equal to a CPHASE gate with the
controlled phase $\Phi$ as defined in Eq.~(\ref{Phi}).

The entangling capability of $U_\mathrm{eff}$ can be assessed by
calculating the concurrence ${\cal C}$ of the states formed by
acting with $U_\mathrm{eff}$ on a separable state of
qubits 1 and 2, and averaging over all such inputs.  This we find to be $ \overline{{\cal C}} = 1/8
  \left|
    15
    -8 \cos (\Phi/2) \cos(2\Theta)
    -7 \cos (4\Theta)
    \right|^{1/2}
  $,
which depends only on the angles $\Phi$ and $\Theta$.  The average
concurrence is thus bounded by $
  \frac{1}{2} \left|\sin(\Phi/4) \right|
  \le
  \overline{{\cal C}}
  \le
  \sqrt{11/32}
  \approx 0.586,
$ where the lower bound comes from the $\Theta=0$ case.  In the
example that we consider later (operation 1 of Table~\ref{tabegp}.),
the off-diagonal angle is $\Theta = 0.42$ and the average
concurrence is $\overline{{\cal C}} = 0.50$.

We will quantify the usefulness of gate $U_\mathrm{eff}$ for QC
purposes solely through the CPHASE angle $\Phi$, even though the gate $U$ really requires two angles $\Phi$ and $\Theta$ to specify its entangling action completely. This one-parameter characterisation is appropriate because, from the foregoing, we know that knowledge of $\Phi$ sets a lower bound on the average entangement generated by the gate. Furthermore, via the construction of Eq.~(\ref{CWCW}), we know explicitly how to form a full CPHASE gate independent of the value of $\Theta$, and indeed independent of the knowledge of its value.
Clearly for $U_\mathrm{eff}$ to be an efficient gate we
require $\Phi$ to be significant.  A value of $\Phi=\pi$ gives the
composite operation as a controlled-$Z$ operation, and in which case
the constituent $U_\mathrm{eff}$ operator may be said to be `as
efficient' as a $\sqrt{\mathrm{SWAP}}$ gate, since two applications
of $\sqrt{\mathrm{SWAP}}$ are also required to form the
controlled-$Z$ operation \cite{los98}.
Experimental demonstration of $U_\mathrm{eff}$ with $\Phi=\pi$
would, in this sense, be equivalent to the demonstration of a
$\sqrt{\mathrm{SWAP}}$ gate.

The question then becomes whether we can obtain a gate
$U_\mathrm{eff}$ with a value of $\Phi=\pi$ within the time
constraints set by $T_Q$ and $T_\mathrm{Rel}$.  The answer is that
if we insist that the evolution of the system be strictly adiabatic, then
for a typical tunneling of $\tau_{34}=0.05$~meV, such a gate takes
several hundred picoseconds.  At this coupling, we have $T_Q
=1.3$~ns, and so the adiabatic pulse time $T$ is a significant
fraction of $T_Q$, which is undesirable.  However, if we allow a
degree of non-adiabaticity in the time evolution, we can
significantly reduce this operation time, as we now discuss.

If we populate the exciton levels then we require that the
population is returned to the qubit sector at the end of the
operation.  We quantify the degree to which this happens with the
average norm on the qubit space.
Let us write as $U_\mathrm{total}$ the time-evolution operator of
the total system. The corresponding operator for the qubit space is
$U = P U_\mathrm{total} P$, where $P$ is the projector onto the 4x4
space. We can then define the average norm on the qubit space  as $
  N_0 =\frac{1}{4} \sum_{i,j=1}^4 |U_{ij}|^2
$.  A perfect return has $N_0=1$ and in this case $U$ necessarily
has the same form as Eq.~(\ref{Ueff}), given the Hamiltonian of Eq.~(\ref{Hblock}).

We then numerically search the space of operations possible within the system for valid gate operations.
For a given tunneling strength $\tau_{34}$, pulse duration
$T$, amplitude $A$, and detuning $\delta$, we numerically integrate
the time-dependent Schr\"{o}dinger equation for the Hamiltonian of Eq.~(\ref{Hblock}) to obtain the systems
evolution from the four initial qubit states $\downarrow
\downarrow$, $\downarrow \uparrow$, $\uparrow \downarrow$ and
$\uparrow \uparrow$.  From these results we can construct $U$, the
evolution operator on the qubit space.  We consider that the
time-evolution can be construed as a valid gate if $N_0 \ge 0.99$.

For a fixed pulse duration $T$, we obtain a range of valid gate operations with various different values of $A$ and $\delta$.  From this set, we are interested in the gate with the most significant gating action which, in terms of the CPHASE angle described above, is the gate for which $|\Phi|$ is closest to $\pi$.
We denote the CPHASE angle of the gate which fullfills the criteria at a given $T$  as $\Phi_\mathrm{opt}$.

\begin{figure}[t]
  \begin{center}
    \epsfig{file=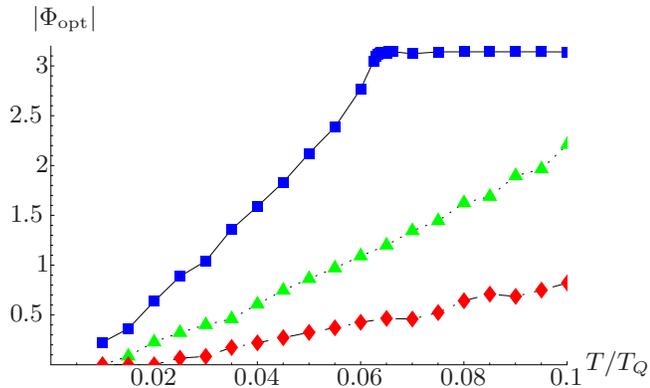,width=1\columnwidth}
    \caption{
    (colour online)
    The absolute value of the optimal CPHASE angle $\Phi_\mathrm{opt}$
    as a function of $T/T_Q$, the pulse duration scaled by
    the qubit isolation time.  Results are shown for three different tunnel
    strengths $\tau_{34}=0.05$~meV (blue squares),
    $\tau_{34}=0.1$~meV(green triangles), and
    $\tau_{34}=0.2$~meV (red diamonds).  As is evident, a phase of $\Phi= \pi$ can
    be obtained at $\tau_{34}=0.05$~meV with $T/T_Q \gtrsim 0.064$.
    In real time units this equates to $ T \approx 84$~ps.
    This value of $\Phi$ can also be achieved with higher tunneling
    amplitudes,
    but this requires longer pulses.
      \label{figmaxP}
    }
  \end{center}
\end{figure}

In Fig. \ref{figmaxP} we plot the optimal angle $\Phi_\mathrm{opt}$
as a function of pulse duration $T$ for several different tunnel
amplitudes $\tau_{34}$.
Let us consider the $\tau_{34}=0.05$~meV case. The angle
$\Phi_\mathrm{opt}$ reaches  a value of $\pi$ at a value of $T/T_Q
=0.064$ and then saturates.  At this coupling, $T_Q=1.3$~ns and so
the pulse duration is $T=84$~ps, which is short compared with both
$T_Q$ and $T_\mathrm{rel}$. For the higher couplings shown in
Fig.~\ref{figmaxP}, we obtain $\Phi_\mathrm{opt}=\pi$ only for
ratios $T/T_Q>0.1$.

In Table \ref{tabegp}, we give parameters and results for two
example $\tau_{34}=0.05$~meV operations: the first corresponds to
the minimum pulse time $T$ required to obtain $\Phi=\pi$, and the second is a
longer duration pulse with higher norm $N_0$.
In the course of the evolution with the first set of parameters, the
exciton states are occupied with a peak population of the order of
40\%.  For initial conditions of three of the four basis states
virtually no population is left in the trion levels at the
conclusion of this operation. However, evolution from $\downarrow
\uparrow$ does leave a small remnant and this reduces the qubit norm
to its value of $N_0=0.991$.  For longer pulse durations, as in the
second operation of Tab. \ref{tabegp}, this remnant can be
eliminated.

Permitting the system to evolve non-adiabatically then improves the
performance of the  gate by reducing the time required to perform a
useful rotation.

\section{Hole-Mixing and Spontaneous Emission}

Other than the direct and kinetic exchanges between the qubits
discussed above, there are two further deleterious processes that we
now discuss: hole-mixing and spontaneous emission.
As we know from the Luttinger-Kohn model of semiconductor bands
\cite{lut55,bro85}, hole-mixing means that, rather than the `bare'
heavy-hole states $\ket{\sfrac{3}{2},\pm\sfrac{3}{2}}$, the actual
hole states are better approximated by
\beq
  \ket{H^{\pm}} &=&
  \cos \theta \ket{\sfrac{3}{2},\pm\sfrac{3}{2}}
  - \sin \theta e^{\mp i\phi}\ket{\sfrac{3}{2},\mp\sfrac{1}{2}}
  \label{Hpm}
  ,
\eeq
where $\ket{\sfrac{3}{2},\mp\sfrac{1}{2}}$ are light-hole states,
and $\theta_m$, $\phi_m$ are mixing angles.
With this mixing, the light-matter interaction Hamiltonian becomes
\beq
  H_{\mathrm{int}} &=& \Omega(t) e^{-i \omega t}
  \left\{\mbox{}
    \cos\theta_m
    e^\dag_{3 \downarrow} h^\dag_{3 H^+}
  \right.
  \nonumber\\
  &&~~~
  \left.
    - \sqrt{1/3} e^{i\phi_m}
    \sin\theta_m
    e^\dag_{3 \uparrow} h^\dag_{3 H^-}
  \right\}
  + \mathrm{H.c.}
  ,
\eeq
where the factor of $\sqrt{1/3}$ in the second term comes from the
different weights of in-plane components of the valence-band wave
functions \cite{bro85,ce06sqR}.

%
\begin{table}[tb]
  \begin{tabular}{|c|c|c|c|c|c|c|c|c|}
    \hline
    ~ & $\tau_{34}$ & $T$ & $T/T_Q$ & $\delta$ & $A$  
      & $\Phi$ &  $N_0$ & ${\cal F}_\mathrm{Rel}$\\
    \hline
     1 & 0.05 & 84  & 0.064 & 0.0729 & 0.10983  & 3.1415   & 0.991 & 0.970\\
     2 & 0.05 & 125 & 0.095 & 0.084 & 0.078025  & 3.1415  & 0.99993 & 0.984 \\
    \hline
  \end{tabular}
  \caption{
    Parameters and results for two example operations.
    $\tau_{34}$ is the tunneling amplitude between dots, and $A$, $T$, and $\delta$
    are the amplitude, duration and detuning of the laser pulse.
    The units of $\tau_{34}$, $\delta$ and
    $A$ are meV, and $T$ is measured in ps.
    For both operations $\Phi \approx \pi$, as require.  Without
    spontaneous emission the fidelity is equal to the norm on the
    qubit space ${\cal F}=N_0$ and this can be made very close to
    unity, as in the second operation.  In the presence of spontaneous emission,
    the fidelity ${\cal F}_\mathrm{rel}$ is slightly reduced.
    \label{tabegp}
  }
\end{table}

This interaction requires that we consider an extra excited-state
sector with $\ket{H^-}$ instead of $\ket{H^+}$. Taking both sectors
into account, we can readily derive an effective Hamiltonian
in the presence of hole mixing.  We find
\beq
  {\cal H}_{\mathrm{mix}} = \cos^2\theta_m {\cal H}_\mathrm{eff}
  + \frac{1}{3}\sin^2\theta_m {\cal H}_\mathrm{-}
  ,
\eeq
where ${\cal H}_\mathrm{eff}$ is as in Eq.~(\ref{Heff})
and
\beq
  {\cal H}_{-}
  &=&
 -\frac{ \Omega^2}{\delta}
 \rb{
   \begin{array}{cccc}
     x_3 & 0 & 0 & 0 \\
     0 & x_3 & y & 0 \\
     0 & y   & x_2 & 0 \\
     0 & 0 & 0 & x_1
   \end{array}
  }
  ,
  \label{eHm}
\eeq
with terms as in Eq.~(\ref{xy}).  Note that the hole-mixing does not
introduce any new matrix elements interactions between qubit states
that were not already present in the non-hole-mixing Hamiltonian.
From this and the symmetry of ${\cal H}_{\mathrm{mix}}$ it
immediately follows that the CPHASE angle is
$
  \Phi_\mathrm{mix} =
  \left\{
    \cos^2\theta_m
    + \frac{1}{3}\sin^2\theta_m
  \right\}\Phi
$, where $\Phi$ is the original angle of Eq.~(\ref{Phi}).  Thus, the
effect of hole-mixing is to slightly decrease the CPHASE angle
$\Phi$, and this can simply be compensated for by a proportionate
increase in Rabi frequency $\Omega$.

We now consider spontaneous emission.  Given our laser excitation
and the magnitudes of the indirect oscillator strengths, the
dominant spontaneous emission path is the direct recombination of
exciton pair $e^\dag_{3\downarrow} h^\dag_{3\Uparrow}$.  As noted
previously, the time-scale of this process is of the order of
$T_\mathrm{rel} =1$~ns.
We can assess the effects through the numerical integration of the
master equation for the system in the Lindblad form.

We judge the quality of the operation in the presence of spontaneous
emission through the fidelity \cite{poy97, che04}, defined as
\beq
  {\cal F} =
  \overline{\ew{\Psi_\mathrm{in} |\widetilde{U}^\dag \rho_\mathrm{out}
  \widetilde{U}|\Psi_\mathrm{in}}}
  \label{fid}
  ,
\eeq
where the overline represents an average over all input states
$\ket{\Psi_\mathrm{in}}$, $\rho_\mathrm{out}$ is the output density
matrix given $\ket{\Psi_\mathrm{in}}$, and $\widetilde{U}$ is the
ideal gate operation.   We define $\widetilde{U}$ by taking the gate
operator without spontaneous emission and renormalising it such that
$N_0=1$. Correspondingly, the fidelity of the operation $U$ without
spontaneous emission is simply given by ${\cal F}=N_0$.

The results of these calculations can be appreciated by again
studying the two operations listed in Table \ref{tabegp}. In this
table we give the fidelities in the presence of spontaneous
emission.  For the first, shorter pulse, spontaneous emission
reduces the fidelity from 0.99 to 0.97. For the longer pulse,
fidelity is reduced from very close to unity to 0.984. This final
value is typical for operations with spontaneous emission, with the
final fidelity lying in the 98---99\% range.

This reduction in fidelity arises largely from population being left
in the trion levels, as attested by the fact that the norm on the
qubit space is affected to a similar extent.  This residual
population occurs because the spontaneous emission removes the
direct exciton $e_3h_3$ but leaves the indirect exciton $e_4 h_3$
untouched.  Since, after the laser pulse has passed, this indirect
state has no route back to the ground state, it is effectively
trapped. This population eventually returns to the ground state,
through either interdot tunneling plus direct spontaneous emission
or, to a lesser extent, indirect emission, but this occurs on a time
scale greater than that considered here. This trapping effect,
although ever present, is small enough that operations of sufficient
quality can still be obtained.

\section{Single-qubit rotations}

An important feature of the current set-up is that it is compatible
with single-qubit operations on the electron spins.  As described in
Ref. \cite{ce06sqR} (see also Ref. \cite{che04}), single qubit
operations can be performed through the excitation of a exciton in
the ground-state levels of a single dot (e.g. the levels 1 in dot
A). The addressability of the individual dots is assured here by
their different Darwin radii.
To obtain arbitrary rotations, a static magnetic field is required in the
$x$-direction (perpendicular to the growth direction),
and it is important that the two-qubit operations function correctly
in the presence of this field.

In InAs dots, both the electron and hole $g$-factors are finite \cite{ce06init,xia06},
and this leads to a
splitting of the single-dot trion levels and the linear polarization
of the optical transitions \cite{ce06init,ce06sqR}.
Assuming that both $g$-factors have the same sign, we can perform
our two-qubit operations with a $\mathbf{H}$-polarized laser (as
defined in Refs.~\cite{ce06init,ce06sqR}) rather than a
$\svec_+$-circularly polarized one. We then derive an effective
Hamiltonian for this situation and find
\beq
  {\cal H}_B = \frac{1}{2}
  \left\{
    {\cal H}_\mathrm{eff}
    + {\cal H }_-\rb{\delta + \Sigma_B}
  \right\}
  \label{H1qbR}
  .
\eeq
where the second term is the same as Eq.~(\ref{eHm}) but with the
detuning $\delta$ replaced by $\delta+\Sigma_B$ throughout.  Here
$\Sigma_B = (g_x^e + g_x^h) \mu_B B_x$, with $g_x^e$ and $g_x^h$ the
electron and hole in-plane $g$-factors, $B_x$ the in-plane field,
and $\mu_B$ the Bohr magneton.
As with the hole-mixing, we see that the magnetic field does not add
any new matrix elements between qubit states, and
thus the operation has the same character as before. Now, however, the CPHASE
angle is given by $\Phi_B = 1/2\left\{\Phi +
\Phi(\delta+\Sigma_B)\right\}$, where the second term is the same as
Eq.~(\ref{Phi}) but with $\delta+\Sigma_B$ replacing $\delta$
throughout.  The contribution of this term is expected to be small
since, with $g$-factors $|g_x^e| = 0.46$ and $|g_x^h|=0.29$
\cite{ce06init,xia06}, at $B=8$~T, we have $|\Sigma_B| \approx
0.4$~meV which is larger than typical values of $\delta$.
In any case, these changes to $\Phi$ can easily be compensated for by
a trivial change in amplitude of the laser.

\section{Conclusions}

We have described a mechanism for obtaining optical-controlled
quantum gates in quantum dots. This may be thought of as a confined
ORKKY interaction in which the exchange between the two qubit spins
is mediated by a set of exciton states delocalized over the double
dot structure.

For this gating to function it is imperative that the two dots be
spaced an appropriate distance apart such that the tunnel coupling
between them is small enough that the kinetic exchange can be
neglected over the course of a gate operation, but large enough that
laser pulses short enough compared with exciton relaxation time can
be used.  Our calculations show that a distance of $d\approx 8$~nm
would be ideal. In this case, significant gates (equivalent to
$\sqrt{\mathrm{SWAP}}$) can be obtained at high fidelities with
pulse durations of the order of 100ps.

The speed of this operation is essentially limited by the occurrence
of the kinetic exchange.  Without this effect, the dots could be
closer together, the tunneling matrix elements larger, and hence the
strength of the gating interaction much stronger.
However, this exchange is unavoidable in vertically-stacked QDs,
since such dots only grow in a stack when the separation between the
two dots is small \cite{xie95}.
This problem, therefore, is not unique to our gating mechanism, but
rather a feature of coupled QDs, and any proposal seeking to show
controlled quantum operations in such a structure must take it into
account.

In this respect, horizontally-coupled quantum dots have an
advantage since, as such dots are coupled in the $xy$ plane, the overlap of
the exited states is a factor of $\sqrt{2}$ greater than that of the
ground states. This means that the strength of the kinetic
exchange interaction is reduced by a factor of one-half relative to
exciton-tunneling.  This in turn means that pulse-durations half
that described here can be used to obtain the same results.

We have also demonstrated that our two-qubit operations function
equally as well with an in-plane magnetic field.  This is important
because it shows that our set-up is compatible the single-qubit
rotations of Ref.~\cite{ce06sqR}.
In combination, this means that we have demonstrated that the laser
excitation of excitonic states provides a realistic,
experimentally-accessible protocol for the fast performance of a
universal set of quantum operations on two electron-spin-qubits in a
coupled double quantum dot.

This work was supported by ARO/NSA-LPS.

\appendix

\section{Single particle wave functions \label{Awfn}}

Our single-particle variational wave functions are separable into
$z$ and $xy$ components.  In the $z$ direction, the wave functions
are those of ground state of a square well of depth $V_0$ and width
$2a$:
\beq
  \chi(z) =
  \left\{
    \begin{array}{cc}
    C \exp( \beta z) & z<-a ,\\
    B \cos(\alpha z) & |z| \le a,\\
    C \exp(-\beta z) & z>a.
    \end{array}
  \right.
\eeq
where
\beq
  \alpha &=& \sqrt{\frac{2 m_e E_z}{\hbar^2}}
  ,
  \\
  \beta &=& \sqrt{\frac{2 m_e (V_0-E_z)}{\hbar^2}}
  .
\eeq
The ground-state energy $E_z$ may be found through solution of the
equation
\beq
  \alpha \tan (\alpha a) = \beta
  .
\eeq

In the $xy$ plane we have the ground-state and first excited-state
Fock-Darwin orbitals
\beq
  \eta_{s}(\omega;\rho, \theta)
    &=& \sqrt{\frac{\omega}{\pi}} e^{-\frac{1}{2} \omega \rho^2}
    ,
  \\
  \eta_{p\pm}(\omega;\rho, \theta)
    &=&  \frac{\omega}{\sqrt{\pi}}\, \rho
     e^{-\frac{1}{2} \omega \rho^2} e^{\mp i \theta},
\eeq
with $\lambda=m \omega/\hbar$. Since the two $p$-orbitals are
degenerate, and angular momentum is conserved, we only need consider
a single orbital and choose $\eta_p=\eta_{p+}$ for concreteness.

The four complete wave functions $\phi_i(\mathbf{r})$ are given in
Eq.~(\ref{4phi}), but these are nonorthogonal.  As our variational
basis we therefore use
\beq
  \psi _1({\bf r}) &=& {\gamma _1}^{-1}
    \left\{\phi _1({\bf r}) - g_1 \phi_2 ({\bf r})\right\}
    ,
  \nonumber\\
 \psi _2({\bf r}) &=& {\gamma _1}^{-1}
    \left\{\phi _2({\bf r}) - g_1 \phi_1 ({\bf r})\right\}
    ,
  \nonumber\\
 \psi _3({\bf r}) &=& {\gamma _3}^{-1}
    \left\{\phi _3({\bf r}) - g_3 \phi_4 ({\bf r})\right\}
    ,
  \nonumber\\
 \psi _4({\bf r}) &=& {\gamma _3}^{-1}
   \left\{\phi _4({\bf r}) - g_3 \phi_3 ({\bf r})\right\}
   ,
\eeq
which are all orthogonal. Here, the overlaps $S_{ij}$ are defined $
  S _{ij} = \int d^3{\bf r}~ \phi _i({\bf r}) \phi _j({\bf r})
$, and the normalization coefficients are given by
\beq
  g _1 &=& = \rb{1-\sqrt{1-S _{12}^2}}/S _{12}
  ,
  \nonumber\\
  g _3 &=& = \rb{1-\sqrt{1-S _{34}^2}}/S _{34}
  ,
  \nonumber\\
  \gamma_1 &=& \sqrt{1 - 2  g_1 S_{12} + g_1^2}
  ,
  \nonumber\\
  \gamma_3 &=& \sqrt{1 - 2 g_3 S_{34}  + g_3^2}
  .
\eeq
%


\end{document}